\documentclass[a4paper,twocolumn,10pt]{article}
\usepackage[top=1.1cm, bottom=2.3cm, left=1.4cm, right=1.4cm]{geometry}

\usepackage{color}
\definecolor{cobalt}{rgb}{0.0, 0.28, 0.67}
\definecolor{airforceblue}{rgb}{0.36, 0.54, 0.66}
\definecolor{armygreen}{rgb}{0.29, 0.33, 0.13}

\usepackage{graphicx}
\usepackage[authoryear]{natbib}
\usepackage{natbibspacing}
\usepackage{subfigure}
\usepackage{tabularx}
\usepackage{txfonts}
\usepackage{color}
\usepackage[breaklinks,colorlinks=true,citecolor=armygreen,linkcolor=black]{hyperref}
\usepackage[hyphenbreaks]{breakurl}

\usepackage{enumitem}

\newcommand{\mydotfill}{\leaders\hbox to 2pt{\hss.\hss}\hfill\phantom{.}}

\newcommand{\apjs}{Astrophysical Journal Supplement Series}
\newcommand{\apj}{Astrophysical Journal}

\newcommand{\mnras}{Monthly Notices of the Royal Astronomical Society}

\usepackage{fancyhdr,lastpage}
\begin{document}

\twocolumn[{
\begin{center}
\phantom{...}
{\Huge\textbf{The world's largest turbulence simulations}}\\
\vspace{0.5cm}
{\large Contribution to the Leibniz Supercomputing Centre (LRZ) Status Reports Book 2016} \\
\vspace{0.5cm}
\Large {
Christoph~Federrath$^1$, 
Ralf~S.~Klessen$^{2,3}$,
Luigi~Iapichino$^4$, \& Nicolay~J.~Hammer$^4$
}\\\vspace{0.5cm}
{\small
{$^1$Research School of Astronomy and Astrophysics, Australian National University (christoph.federrath@anu.edu.au)}\\
{$^2$Zentrum f\"ur Astronomie der Universit\"at Heidelberg, Institut f\"ur Theoretische Astrophysik (klessen@uni-heidelberg.de)}\\
{$^3$Universit\"at Heidelberg, Interdisziplin\"ares Zentrum f\"ur Wissenschaftliches Rechnen}\\
{$^4$Leibniz-Rechenzentrum der Bayerischen Akademie der Wissenschaften (luigi.iapichino@lrz.de, nicolay.hammer@lrz.de)}\\
}
\vspace{0.5cm}
\end{center}
}]

\section*{\color{airforceblue}Introduction}
Understanding turbulence is critical for a wide range of terrestrial and astrophysical applications. For example, turbulence on earth is responsible for the transport of pollutants in the atmosphere and determines the movement of weather patterns. But turbulence plays a central role in astrophysics as well. For instance, the turbulent motions of gas and dust particles in protostellar disks enables the formation of planets. Moreover, virtually all modern theories of star formation rest on the statistics of turbulence \citep{PadoanEtAl2014}. Especially the theoretical assumptions about turbulence behind star formation theories allow the prediction of star formation rates in the Milky Way and in distant galaxies \citep{FederrathKlessen2012}. Interstellar turbulence shapes the structure of molecular clouds and is a key process in the formation of filaments which are the building blocks of star-forming clouds. 
The key ingredient for all these models is the so-called sonic scale. The sonic scale marks the transition from supersonic to subsonic turbulence and produces a break in the turbulence power spectrum from $E \propto k^{-2}$ to $E \propto k^{-5/3}$. While the power-law slopes of -2 and -5/3 for the supersonic and subsonic parts of the spectrum have been measured independently, there is no simulation currently capable of bridging the gap between both regimes. This is because previous simulations did not have enough resolution to separate the injection scale, the sonic scale and the dissipation scale.

\begin{figure}[hb]
\centerline{\includegraphics[width=0.91\linewidth]{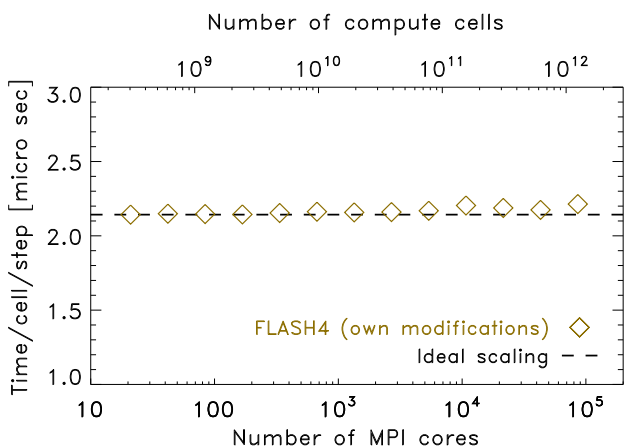}}
\caption{Weak scaling of the customized version of the FLASH code, used during the SuperMUC scale-out workshop on Phase 2 in 2015. The diamonds indicate the scaling tests of the FLASH code, while ideal scaling is represented by the dashed line.}
\label{fig:scaling}
\end{figure}

The aim of this project is to run the first simulation that is sufficiently resolved to measure the exact position of the sonic scale and the transition region from supersonic to subsonic turbulence. A simulation with the unprecedented resolution of $10000^3$ grid cells will be needed for resolving the transition scale.

\begin{figure}
\centerline{\includegraphics[width=0.91\linewidth]{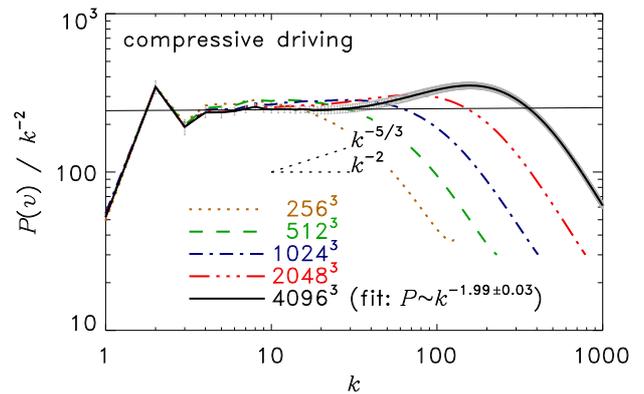}}
\caption{Power spectrum from highly-compressible, supersonic turbulence simulations (compressive driving), demonstrating a $k^{-2}$ scaling \citep{Federrath2013}.}
\label{fig:spectrum}
\end{figure}

\section*{\color{airforceblue}Results}

\begin{figure*}[htb]
\centerline{\includegraphics[width=1.0\linewidth]{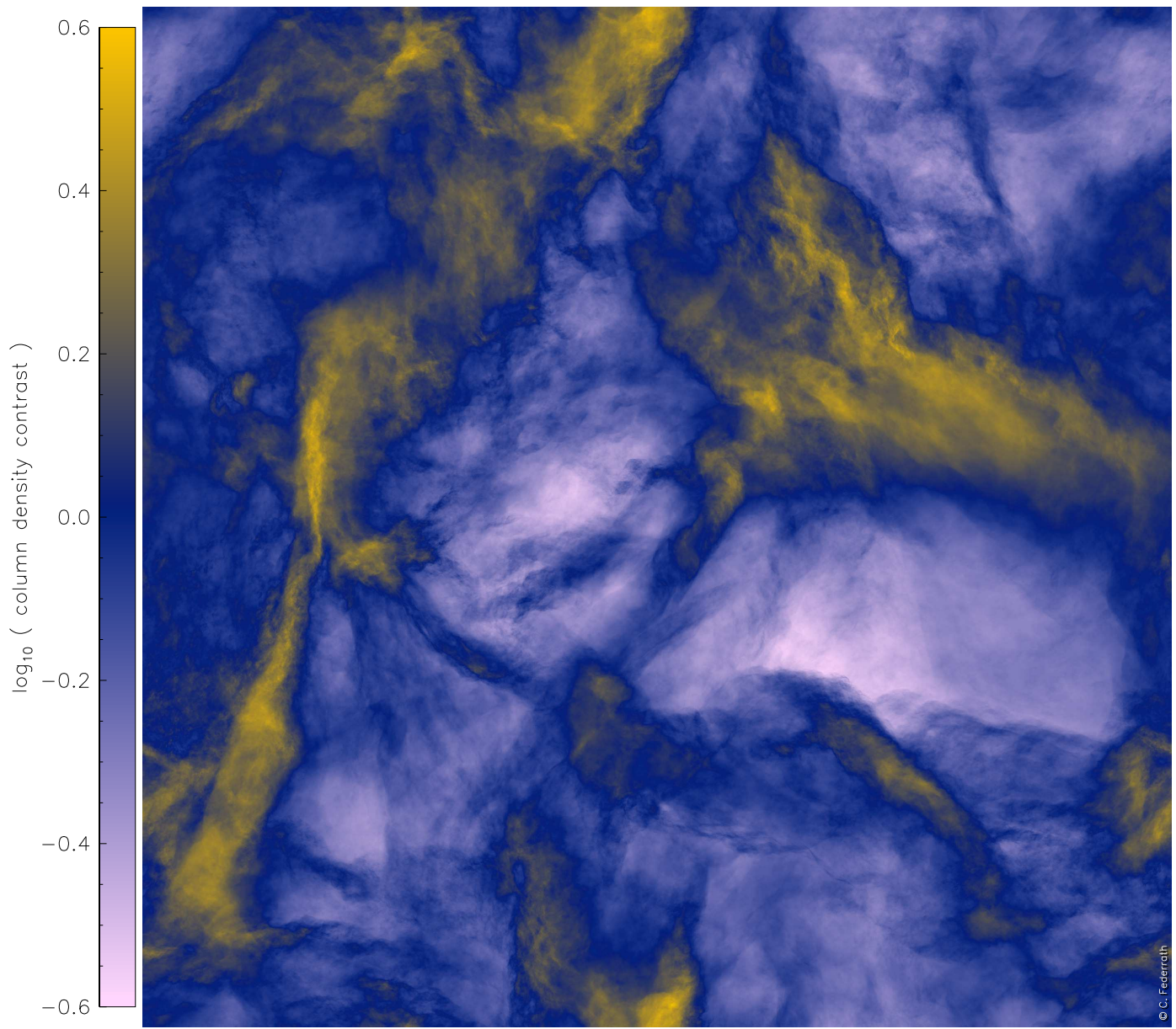}}
\caption{Column gas density projection in our simulation of supersonic turbulence with a grid resolution of $10048^3$ cells (Federrath et al., in preparation).}
\label{fig:snaphots}
\end{figure*}

In the framework of a GAUSS Large Scale Project, an allocation exceeding 40~million core-h has been granted to this project on SuperMUC. The application used for this project is FLASH, a public, modular grid-based hydrodynamical code for the simulation of astrophysical flows \citep{FryxellEtAl2000}. The parallelisation is based entirely on MPI. In the framework of the SuperMUC Phase 2 scale-out, the current code version (FLASH4) has been optimised to reduce the memory and MPI communication requirements. In particular, non-critical operations are now performed in single precision, without causing any significant impact on the accuracy of the results. In this way, the code runs with a factor of 4.1 less memory and 3.6 times faster than the version used for the previous large-scale project at LRZ \citep{Federrath2013}, and scales remarkably well up to the full machine on SuperMUC Phase 2 (see Figure~\ref{fig:scaling}).

Our current $10048^3$ simulation has been nearly completed at the time of writing, and data processing is in progress. Some early impression of the forthcoming results can be seen from the highlights of the work of \citet{Federrath2013}, based on the previous large-scale project on turbulence simulations (up to $4096^3$ grid cells), selected as the SAO/NASA ADS paper of the year 2013.

Highly-compressible supersonic turbulence is complex, if compared to the subsonic, incompressible regime, because the gas density can vary by several orders of magnitude. Using three-dimensional simulations, we have determined the power spectrum in this regime (see Figure~\ref{fig:spectrum}), and found $E \propto k^{-2}$, confirming earlier indications obtained with much lower resolution \citep{KritsukEtAl2007}. The resolution study in Figure~\ref{fig:spectrum} shows that we would not have been able to identify this scaling at any lower resolution than $4096^3$ cells. Extremely high resolution and compute power are absolutely necessary for the science done here. 

Figure~\ref{fig:snaphots} displays the unprecedented level of detail in density structure achieved with our current $10048^3$ simulation. This visualization highlights the enormous complexity of the turbulent structures on all spatial scales covered in these simulations. Simulation movies are available online (see links below).

\vspace{-0.2cm}
\section*{\color{airforceblue}Future Work}

Turbulence has a wide range of applications in science and engineering, including the amplification of magnetic fields, star and planet formation, mixing of pollutants in the atmosphere, fuel ignition in engines, and many more. Generating the huge dataset of turbulence presented here, we have begun to reach the technical limits of what is feasible on any supercomputer in the world to date. We are currently pushing the boundaries even further by running the world's first turbulence simulation with $10048^3$ grid cells on SuperMUC. We hope to unravel the statistics of supersonic and subsonic, magnetized turbulence in the near future, with cutting-edge supercomputing systems provided by the LRZ.

\section*{\color{airforceblue}References and Links}

\setlength{\bibspacing}{2.5pt}
\renewcommand*{\refname}{\vspace*{-10mm}}


\begin{thebibliography}{5}
\expandafter\ifx\csname natexlab\endcsname\relax\def\natexlab#1{#1}\fi

\bibitem[{{Federrath}(2013)}]{Federrath2013}
{Federrath}, C. 2013, \mnras, 436, 1245

\bibitem[{{Federrath} \& {Klessen}(2012)}]{FederrathKlessen2012}
{Federrath}, C., \& {Klessen}, R.~S. 2012, \apj, 761, 156

\bibitem[{{Fryxell} {et~al.}(2000){Fryxell}, {Olson}, {Ricker}, {Timmes},
  {Zingale}, {Lamb}, {MacNeice}, {Rosner}, {Truran}, \&
  {Tufo}}]{FryxellEtAl2000}
{Fryxell}, B., {Olson}, K., {Ricker}, P., {et~al.} 2000, \apjs, 131, 273

\bibitem[{{Kritsuk} {et~al.}(2007){Kritsuk}, {Norman}, {Padoan}, \&
  {Wagner}}]{KritsukEtAl2007}
{Kritsuk}, A.~G., {Norman}, M.~L., {Padoan}, P., \& {Wagner}, R. 2007, \apj,
  665, 416

\bibitem[{{Padoan} {et~al.}(2014){Padoan}, {Federrath}, {Chabrier}, {Evans},
  {Johnstone}, {J{\o}rgensen}, {McKee}, \& {Nordlund}}]{PadoanEtAl2014}
{Padoan}, P., {Federrath}, C., {Chabrier}, G., {et~al.} 2014, Protostars and
  Planets VI, 77

\end{thebibliography}

\vspace{0.5cm}
\small{\noindent
Gauss Centre for Supercomputing:\\
\url{http://www.gauss-centre.eu/gauss-centre/EN/Projects/Astrophysics/federrath_astrophysics_weltrekord.html?nn=1345700} \\[5pt]
Uni Heidelberg:\\
\url{http://www.ita.uni-heidelberg.de/~chfeder/pubs/supersonic/supersonic.shtml?lang=en}
}
\clearpage

\end{document}